\documentclass{article}[12pt]
\usepackage{graphicx}
\newcommand{\eqref}[1]{(\ref{#1})}
\def\cosec{{\rm cosec\,}}
\begin{document}
\begin{center}
\Large{\bf{Bound States and Band Structure - a Unified Treatment
through the Quantum Hamilton - Jacobi Approach}}\\
\end{center}
\begin{center}
S. Sree Ranjani,\footnote{ranjani@physics.iitm.ac.in}
A. K. Kapoor,\footnote{akksp@uohyd.ernet.in}\\
School of Physics, University of Hyderabad, Hyderabad, 500 046, INDIA.\\
P. K. Panigrahi, \footnote{prasanta@prl.ernet.in}\\
Physical Research Laboratory, Navrangpura, Ahmedabad, 380 009, INDIA.\\
\end{center}
\begin{abstract}
     We analyze the Scarf potential, which exhibits both discrete energy
bound states and energy bands, through the quantum Hamilton-Jacobi
approach. The singularity structure and the boundary conditions in
the above approach, naturally isolate the bound and periodic
states, once the problem is mapped to the zero energy sector of
another quasi-exactly solvable quantum problem. The energy
eigenvalues are obtained without having to solve for the
corresponding eigenfunctions explicitly. We also demonstrate how
to find the eigenfunctions through this method.
\end{abstract}

\section{Introduction}

\indent
     Study of periodic potentials has evoked renewed interest in
the literature in light of their appearance in Bose-Einstein
condensates \cite{br}, \cite{fo} and  photonic crystals
\cite{dim}, \cite{jo}, \cite{fs}, \cite{fs1}. It has been possible
to experimentally change structure of the potential, so as to
produce superfluid - insulator transition \cite{na}, the former
having delocalized states and the latter localized ones. In this
context, traditional Kronig-Penny model \cite{kit} is used for
illustrative purposes, wherein the known wave functions lead to
transcendental equations involving energy and momentum, when
appropriate boundary conditions are implemented. The possibility
of investigating superfluid-insulator type transitions mentioned
above does not arise here, due to the lack of any control
parameter. Quite sometime back, Scarf showed that a solvable model
exists, which exhibits both discrete bound states and band spectra
\cite{3scf}, as a function of the coupling parameter. The group
theoretical aspects of this problem have recently been
investigated \cite{3li}.  The fact that Scarf potential yields
both bound states and band structure, as a function of a coupling
parameter, makes this model an ideal one to study the interplay of
these two types of distinct behavior in a given quantal problem.

 The goal of this paper is to first map the Scarf eigenvalue problem into
 the zero energy sector of  another quasi-exactly solvable (QES) problem. We
 then use the quantum Hamilton-Jacobi (QHJ) approach \cite{lea}, which
 naturally takes advantage of the singularities of the new potential, to
 isolate the domains corresponding to discrete and band spectra. The
subtle aspects of the boundary conditions in quantum mechanics,
which lead to the existence of both bound states and band
structure in the Scarf potential, come out naturally in this
approach. We then proceed to obtain the eigenvalues and
eigenfunctions, for both the cases. In this procedure, the energy
eigenvalues can be obtained, without finding the eigenfunctions.
The QHJ formalism, being formulated in the complex domain where
the non-linear Riccati equation replaces the Schr\"odinger
equation, makes use of powerful theorems in complex variable
theory to obtain the solutions.

      Apart from the fact that QHJ formalism is relatively
new and requires detailed study, this approach may provide a
different perturbative treatment for the traditional problems. As
will be clear form the text, WKB approximation scheme is close to
this method \cite{wkb}. In the following section, we briefly
describe the working principles of the QHJ formalism, which is
then used for the analysis of the Scarf potential in Section 3.
The  origin of the bound and the band spectra is then illustrated,
without  getting into the explicit computation of the eigenvalues,
whose details are given in Section 3.2. We obtain the solutions
pertaining to both the spectra. We conclude in the final section
after pointing out various directions for future investigations.

\section{Quantum Hamilton - Jacobi formalism}

         The QHJ formalism, formulated as a theory
analogous to the classical canonical transformation theory
\cite{d2}, \cite{jor}, \cite{sw}, was proposed by Leacock and
Padgett in 1983. It has been applied to one dimensional bound
state problems and separable problems in higher dimensions
\cite{lea}. In our earlier studies, we have shown that one could
use the QHJ formalism to analyze one dimensional exactly solvable
(ES), quasi - exactly solvable models, consisting of both periodic
and aperiodic potentials \cite{3sree}, \cite{geo}, \cite{geoth},
\cite{akk}, \cite{ppqes}, \cite{sreeth} and the recently
discovered PT symmetric potentials \cite{pt}. The advantage of
this method lies in the fact that it requires a modest
understanding of basic quantum mechanics and complex analysis as a
prerequisite.

        In this formalism,  the logarithmic derivative
of the wave function $\psi(x)$, given by
\begin{equation}
p= -i\hbar \frac{d}{dx}\ln \psi(x),     \label{e1}
\end{equation}
plays an important role. This is referred to as the quantum
momentum function (QMF), since it is defined analogous to the
classical momentum function as, $p = \frac{dS}{dx}$. Here, $S$ is
the Hamilton's characteristic function which is related to the
wave function by $\psi(x) = \exp(iS/\hbar)$. Substituting
$\psi(x)$ in terms of $S$ in the Schr\"odinger equation,
\begin{equation}
-\frac{\hbar^2}{2m}\frac{d^2\psi(x)}{dx^2} + V(x) \psi(x) =
E\psi(x),
\end{equation}
 and using the relation between $p$ and $\psi$, one obtains the non-linear Riccati
 equation:
\begin{equation}
p^2 - i \hbar p^{\prime} =  2m(E - V(x)).  \label{e3}
\end{equation}
The above equation is known as the QHJ
equation; here $x$ is treated as a complex variable, thereby
extending the definition of $p$ to the complex plane. We show that
one can arrive at the required results by studying the singularity
structure of the QMF.\\
 \noindent
{\bf Singularity structure}\\
The QMF has two types of singularities, the moving and the fixed
singularities. From \eqref{e1} one can see that, the $n$ nodes of
the $n^{th}$ excited state, whose locations depend on the initial
conditions and  energy, correspond to the singularities of $p$.
These are known as the moving singularities. It is a fact that
only poles can appear as moving singularities in the solutions of
the Riccati equation. One can calculate the residue at a moving
pole $x_0$, where $V(x)$ is analytic, by doing a Laurent expansion
of $p$ around $x_0$ as,
\begin{equation}
p = \sum_{k=1}^{l} (x-x_0)^{-k} + \sum_{k=0}^{\infty}(x-x_0)^k.
\label{e4}
\end{equation}
Substituting this in \eqref{e3} and  comparing individually the
coefficients of different powers of $x-x_0$, one obtains $l = 1$,
with the corresponding residue equalling  $-i\hbar$.

    The fixed singularities originate from the potential and are
present in all the solutions of the Riccati equation.  One can
calculate the residue at the fixed poles in the same way, as is
done for the moving poles. Owing to the quadratic nature of the
QHJ equation, one obtains two solutions. In order to arrive at the
right solution, one needs to choose the residue that gives the
correct physical behavior. The right value of the residue is
chosen by applying the appropriate boundary conditions, details of
which will be given in the text, as and when required. Thus,
knowing the singularity structure of  QMF and the behavior of $p$
at infinity, one gets the complete form of the QMF. In all the
models studied so far, including the periodic potentials, the
assumption that, {\it the QMF has finite number of singularities,
is equivalent to saying that the point at infinity is an isolated
singular point}, has been found to be true. We expect it to be
valid for the present case also.

     For most exactly solvable models, the QMF has been found to
be a rational function. As is known,  for a rational function the
sum of all  residues including that at infinity is zero. This
result has been used to obtain the energy eigenvalues for all the
models studied in the QHJ approach. It should be pointed out that,
this condition is equivalent to the  exact quantization condition
satisfied by the action $J$ \cite{lea}:
\begin{equation}
J = \oint_C pdx = n \hbar.   \label{qc}
\end{equation}
Hence, for the case of Scarf potential, one first tries to bring
the QMF into a rational form through a suitable change of
variable, as discussed in the next section. It is interesting to
note that in the classical limit,
\begin{equation}
p \rightarrow p_c = \sqrt{2m(E - V(x))},  \label{bc}
\end{equation}
where $p_c$ is the classical momentum. The QHJ quantization
condition then leads to the WKB approximation scheme. The boundary
condition \eqref{bc} was originally used by Leacock and Padgett
to obtain the constraints on the residues\cite{lea}.
\section{The Scarf Potential}
    The Scarf potential is given by
\begin{equation}
V(x) = - \left(\frac{(\frac{1}{4} -s^2)\pi^2}{2ma^2
\sin^2(\frac{\pi x}{a})}\right),    \label{3es0}
\end{equation}
where, $a$ is the potential period. One finds that in the range
$s>1/2$, the potential is an array of infinite potential wells as
shown in Fig 1.

\begin{figure}
\centering
\includegraphics[width=3in]{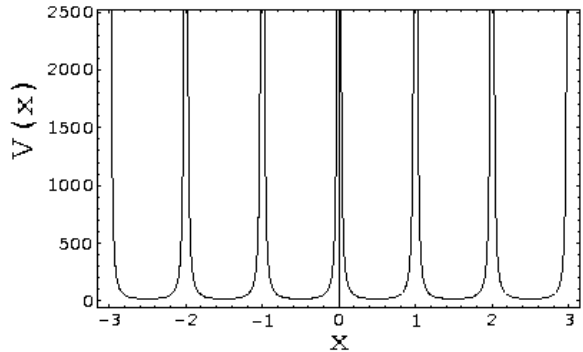}
\caption{Scarf potential, with $s=2$, $a=1$, allows bound states.}
\end{figure} A quantum particle is then confined to only one
well, implying
that the wave function should vanish at $x = \pm a$. Thus, in the
above range, the potential exhibits bound state spectra. As shown
in Fig.2, in the range $0 < s< 1/2$, the potential is similar to
that of a potential in a crystal lattice,
\begin{figure}
\centering
\includegraphics[width=3in]{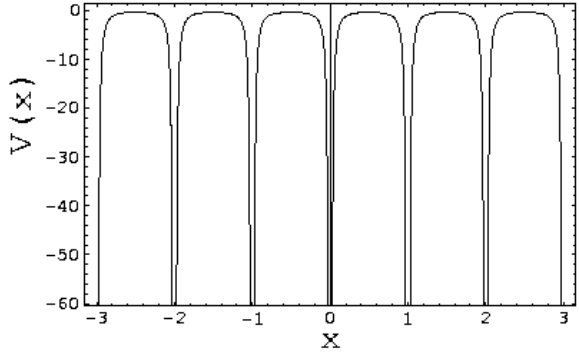}
\caption{Scarf potential, with $s=0.4$, $a=1$, allows band
structure. }
\end{figure}

leading to the possibility of energy bands. In this scenario, a
particle can escape to infinity. Therefore the wave function need
not vanish at $x = \pm a$. However, $\psi(x)$ should not diverge
anywhere, on physical grounds.

    The QHJ equation for the Scarf potential, with $p=-iq$ and $\hbar
=1$ in \eqref{e3}, is given by,
\begin{equation}
q^2 + q^{\prime}+\frac{\pi^2 }{a^2} \left(\lambda^2
+\frac{(\frac{1}{4}-s^2)}{\sin ^2(\frac{\pi     x}{a})}\right) =
0, \label{3es1}
\end{equation}
where $\lambda^2  = 2mEa^2/\pi^2$. We perform a change
of variable using
\begin{equation}
 y =  \cot \left(\frac{\pi x}{a}\right),  \label{3es1a}
\end{equation}
which transforms \eqref{3es1} to
\begin{equation}
q^2(y) - \frac{\pi}{a}(1+y^2)\frac{dq}{dy} + \frac{\pi^2 }{a^2}
\left(\lambda^2 +(\frac{1}{4}-s^2)(1+y^2)\right) = 0.  \label{e10}
\end{equation}
In order to get all the coefficients in the above equation to a
rational form, which in turn will easily yield the
singularities of the QMF, we use the transformation equations
\begin{equation}
q = -\frac{\pi \hbar }{a}(1+y^2) \phi  \,\,; \,\,\, \phi = \chi
-\frac{y}{1+y^2}. \label{3es3}
\end{equation}
This leads to the  QHJ equation in terms of $\chi$ as,
\begin{equation}
\chi^2 +\frac{d \chi}{dy} + \frac{\lambda^2 -1} {(y^2 +1)^2} +
\frac{(\frac{1}{4} -s^2)}{y^2+1} = 0.    \label{3es4}
\end{equation}
Henceforth, the above equation will be treated as the QHJ equation
and $\chi$ as the QMF. It is interesting to note that,
substituting  $\chi=\frac{d}{dy}(\ln(\tilde{\psi}(y))$ in the
above equation, one gets a Schr\"odinger equation, which describes
the zero energy sector of the potential,
$(\lambda^2 -1)/(y^2 +1)^2 +(\frac{1}{4}-s^2)/(y^2+1)$. By
analyzing the singularity structure of this quasi-exactly solvable
problem, we obtain the required results for the solvable Scarf
potential, as shown below.
\subsection{Form  of the QMF \protect \( \chi  \protect \) }
     The QMF has $n$ moving poles with residue
one on the real line, as is clear from the Riccati equation. From
\eqref{3es4}, one can see that $\chi$ has fixed poles at $y = \pm
i$. Making use of the assumption that the QMF has finite number of
moving poles, one can write $\chi$ in the rational form,
separating its analytical and singular parts as
\begin{equation}
\chi = \frac{b_1}{y-i} +\frac{b^{\prime}_1}{y+i}
+\sum_{k=0}^{n}\frac{1}{y-y_k} + Q. \label{3es5a}
\end{equation}
Here, $b_1$ and $b^{\prime}_1$ are the residues at $y =  i$ and $y
=-i$ respectively and the summation term describes the sum of all
the singular parts coming from the moving poles. Note that,
$\sum_{k=0}^{n}\frac{1}{y-y_k} = \frac{P^{\prime}_n(y)}{P_n(y)}$,
where $ P_n(y)$ is an $n^{th}$ degree polynomial. The quantity $Q$
represents the analytic part of $\chi$ and from \eqref{3es4} one
can see that $\chi$ is bounded for large $y$. Thus, from
Liouville's theorem, $Q$ is a constant; denoting it as $C$,
\eqref{3es5a} can be written as,
\begin{equation}
\chi = \frac{b_1}{y-i} +\frac{b^{\prime}_1}{y+i}
+\frac{P^{\prime}_n(y)}{P_n(y)} + C. \label{3es5}
\end{equation}
One can calculate the residues at the fixed poles
$y = \pm i$, by making a Laurent expansion of $\chi$ around the
pole. For example, to calculate the residue at $y=i$, we expand
$\chi$ as,
\begin{equation}
\chi = \frac{b_1}{y-i} + a_0 + a_1(y -i)+ \cdots ~~.
\label{3es5b}
\end{equation}

Comparing the coefficients of different powers of $(y-i)$
individually, one obtains
\begin{equation}
b_1 = \frac{1 \pm \lambda}{2}.     \label{3es6a}
\end{equation}
Similarly the other residue at $y= -i$ is found to be
\begin{equation}
b^{\prime}_1 = \frac{1 \pm \lambda}{2}.  \label{3es6}
\end{equation}
To find the eigenvalues, we now  make use of the fact that, for a
rational function, the sum of all the residues equals zero.
As noted earlier, this is equivalent to the quantization condition
\eqref{qc} of Leacock and Padgett.  Thus, we obtain
\begin{equation}
b_1 +b^{\prime}_1 + n = d_1,        \label{3es7}
\end{equation}
where $d_1$ is the residue at infinity, which is calculated by
taking Laurent expansion of $\chi$ around the point at infinity:
\begin{equation}
\chi = d_0 +\frac{d_1}{y} + \frac{d_2}{y^2} +\cdots ~~.
\label{3es10}
\end{equation}
Substitution of the above in the QHJ equation yields,
\begin{equation}
d^2_1 -d_1 + (\frac{1}{4} -s^2) = 0, \label{3es11}
\end{equation}
from which, the values of $d_1$ can be deduced:
\begin{equation}
  d_1 = \frac{1 \pm 2s}{2}.  \label{3es12}
\end{equation}
Substituting the values of the residues in \eqref{3es7}, one obtains
\begin{equation}
n = - \frac{1}{2} \pm s \mp \lambda,  \label{3es12a}
\end{equation}
which gives the degree of the polynomial $P_n(y)$ in \eqref{3es5}.
From the definition of $\lambda$, one can see that if $E<0$,
$\lambda$ becomes imaginary, in which case \eqref{3es12a} will not
be satisfied. Thus, from the above equation,  we have the
condition $E >0$, which in turn implies $\lambda >0$ and real.
Hence, for any range of $s$, the energy eigenvalues are
greater than zero. With this  condition on $\lambda$, we now
proceed to select the values of the residues at the fixed poles
and at infinity, which will give us the physically acceptable
results.

\subsection{Choice of the residues }
    One needs to use the boundary conditions obeyed by the QMF \cite{lea} to choose the
right value of residues. Although, there are several ways of
implementing the boundary conditions in the QHJ formalism, we have
chosen the one closest to the conventional approach for clarity.
First, we shall fix the value of the residue at infinity. From the
prior discussion of the potential, we know that the wave functions
should not become infinite  anywhere, in particular, for $x=  \pm
a$. From \eqref{e1}, one can obtain $\psi(x)$ in terms of the QMF.
Writing $p=-iq$ and doing the change of variable,  one obtains the
wave function:
\begin{equation}
\psi(y) = \exp \left(-
\int\frac{a}{\pi}\left(\frac{q}{1+y^2}\right)dy\right).
\label{3es13a}
\end{equation}
Using the transformation equations in \eqref{3es3}, the above
expression for the wave function becomes
\begin{equation}
\psi(y) = \exp\left(\int \left(\chi - \frac{y}{1+y^2}\right)dy
\right).   \label{3es13}
\end{equation}
For large $y$, the leading behavior of $\chi$  is obtained as
$\chi \sim \frac{d_1}{y}$, which when substituted in
\eqref{3es13}, yields,
\begin{equation}
\psi(y)  \sim \exp \left( \int \left( \frac{d_1}{y} - \frac{y}{1+y^2}
\right)dy\right)   \label{3es14}
\end{equation}
\begin{equation}
 \sim \frac{y^{d_1}} {( y^2 +1)^{1/2}}. \label{3es15}
\end{equation}
Using the value of $d_1$ from \eqref{3es12} in the above equation,
one obtains
\begin{equation}
\psi(y) \sim  \frac{y^{\frac{1}{2} \pm s}} {(y^2+1)^{1/2}}.
  \label{3es16}
\end{equation}
    For $0 < s < \frac{1}{2}$, one can see that $\psi \rightarrow 0$, in the
limit $ y \rightarrow \infty, x \rightarrow ma$, with $m$ being an
integer, for both the values of $d_1$. This range corresponds to
the case where the potential exhibits band structure.

  For $s>1/2$, $\psi(y) \rightarrow 0$, in the limit
  $ y \rightarrow\infty, x \rightarrow ma$, with $m$ being an integer,
  only if $d_1$ takes the value $ 1/2 -s $. In this way, the two
different ranges of the potential parameter $s$ emerge
simultaneously, while  fixing the values of $d_1$. In order to
select the values of $b_1$ and $b^\prime_1$, we note that the
bound state and band edge wave functions of one dimensional
potentials are non-degenerate and have definite parity. Parity
operation requires that $\chi(-y) = -\chi(y)$, which in turn gives
\begin{equation}
b_1 = b^\prime_1.      \label{3es17}
\end{equation}
With the above constraint, \eqref{3es7} becomes
\begin{equation}
2b_1 +n = d_1.    \label{3es17a}
\end{equation}
     Finiteness of the wave function
as $ x \rightarrow \infty$, gives the values of $d_1$ in the two
ranges as
\begin{equation}
 d_1  =  \left\{ \begin{array}{cc}
         \quad \,\,\,  \frac{1\pm 2s}{2}  \quad\mbox{for}\quad
               0 < s <1/2,
            \\ \quad \mbox{}\quad
    \\ \frac{1-2s}{2}  \quad \mbox{for} \quad s >1/2.
           \end{array}
           \right.        \label{3es17ab}
\end{equation}
From the parity constraint, one obtains the restriction on the values
of the residues at the fixed poles as $b_1 = b^{\prime}_1$. Using
these results, we proceed proceed to calculate the solutions for
the two ranges.

\subsection{Case 1 : Band spectrum}
    In the range $0<s<1/2$, we have seen that $d_1$ can take both the
values of the residues. Taking all the possible
combinations of the residues, with $b_1 = b^{\prime}_1$ and
substituting them in \eqref{3es7}, we evaluate $n$, the degree of
the polynomial $P_n(y)$. There are four combinations
forming four different sets, as given in
the fifth column of table I.
\begin{table}[p]
\begin{center}
 \bf{ Table 1. All possible combinations of residues in the range
$0<s<1/2$.}
\end{center}
 \vskip0.5cm
\begin{center}
\begin{tabular}{|c|c|c|c|c|c|}
\hline
\multicolumn{1}{|c|}{Set}
&\multicolumn{1}{|c|}{$b_1$}
&\multicolumn{1}{|c|}{$b^{\prime}_1$}
&\multicolumn{1}{|c|}{$d_1$}
&\multicolumn{1}{|c|}{$n = d_1 -b_1-b^{\prime}_1$ }
&\multicolumn{1}{|c|}{remark}\\
\hline
& & & & & \\
1 & $\frac{1-\lambda}{2}$ & $\frac{1-\lambda}{2}$ & $d_1 = 1/2 -s$ &
$\lambda -s -\frac{1}{2}$ &  $\lambda > (s +\frac{1}{2})$\\
& & & & &\\
2 & $\frac{1-\lambda}{2}$ & $\frac{1-\lambda}{2}$ & $d_1 = 1/2 +s$ &
$\lambda +s -\frac{1}{2}$ &  $\lambda > -(s - \frac{1}{2})$\\
& & & & & \\
3 & $\frac{1+\lambda}{2}$ & $\frac{1+\lambda}{2}$ & $d_1 = 1/2 -s$
&$-\lambda -s -\frac{1}{2}$ & not valid\\
& & & & &\\
4 & $\frac{1+\lambda}{2}$ & $\frac{1+\lambda}{2}$ & $d_1 = 1/2 +s$&
$-\lambda + s -\frac{1}{2}$ & not valid\\
& & & & &\\
\hline
\end{tabular}

\end{center}
\end{table}
Since $n$ needs to be positive, from table I, we  pick only those
sets which give a positive integral value for $n$. As seen
earlier, $\lambda$ is a positive real constant. Thus, only the
sets 1 and 2 will yield positive values for $n$ and hence; the
other two sets are ruled out. Taking the values of $b_1$ and $d_1$
from the sets 1 and 2, substituting them in \eqref{3es17a} and
using the definition of $\lambda$  and $s$, we obtain the
expressions for the energy eigenvalues corresponding to the two
band edges of the $n^{th}$ band as,
\begin{equation}
E^{\pm}_n = \frac{\pi^2 }{2ma^2}\left( n+\frac{1}{2} \pm s
\right)^2. \label{3es22}
\end{equation}
Here, $E_n^{\pm}$ correspond to the upper and lower band energies
of the $n^{th}$ band. These results match with the solutions given
in \cite{3scf} and \cite{3li}.

    The corresponding wave functions follow from
 \eqref{3es5} and \eqref{3es13}:
\begin{equation}
\psi(y) = (y^2 +1)^{b_1 - \frac{1}{2}}P_n(y).
  \label{3es23}
\end{equation}
To obtain the expression for the polynomial, we substitute $\chi$
from \eqref{3es5} in the QHJ equation, which gives a second order
differential equation:
\begin{eqnarray}
&&P^{\prime\prime}_n(y) +
 \left(\frac{4b_1 y}{y^2+1}\right)P^{\prime}_n(y)\,\,\, + \nonumber \\ & &
   \qquad \left(\frac{1/4 - s^2}{y^2 +1} + \frac{4b_1^2y^2
 +2b_1(1-y^2) +  \lambda^2 -1}{(y^2  +1)^2}\right)P_n(y) = 0.   \label{3es24}
\end{eqnarray}
The sets 1 and 2 have $b_1 =
 (1-\lambda)/2$, which yields
\begin{equation}
P^{\prime\prime}_n(y) + \left(\frac{2(1-\lambda)y}{y^2
 +1}\right)P^{\prime}_n(y) +\frac{\frac{1}{4}-s^2
 +\lambda^2 - \lambda}{(y^2 +1)}P_n(y) = 0.   \label{3es24a}
\end{equation}
From \eqref{3es22}, one can
see that $\lambda$ has two values $\lambda = n \pm s +
\frac{1}{2}$. Substituting these in the above equation, one
obtains two differential equations corresponding to the two energy
eigenvalues $E_n^{\pm}$ :
\begin{equation}
(y^2 +1)P_n^{\prime\prime}(y) + (1 - 2n \mp 2s)y
  P_n^{\prime}(y) + n(n\pm 2s)P_n(y) = 0. \label{3es24b}
\end{equation}
Defining $y=it$, the above equation takes the form of the well known Jacobi
differential equation
\begin{equation}
(1-t^2)P^{\prime\prime}_n(t) + (\nu_1 -\nu_2 -t(\nu_1 +\nu_2
  +2))P^{\prime}_n(t) +n(n +\nu_1 +\nu_2 +1)P_n(t) = 0,  \label{3es23a}
\end{equation}
with $\nu_1 = \nu_2 = -n \mp s -1/2$, for the corresponding two
$\lambda$ values. The expression for the two band edge wave
functions for the $n^{th}$ band are given by,
\begin{equation}
\psi(y) = (y^2 +1)^{-\frac{\lambda}{2}}P^{\nu_1,
\nu_2}_n(-iy),\label{3es20}
\end{equation}
with their respective $\nu_1, \nu_2$ values corresponding to
$\lambda = n \pm s +1/2$.
\subsection{Case 2 : Bound state spectrum}
    We proceed in the same way as in case 1 {\it i.e.},
take all possible combinations of $b_1, b^{\prime}_1$ and $d_1$,
keeping $b_1 =b^{\prime}_1 $ in \eqref{3es17a}. Since  $d_1$ can
take only one value $1/2 -s$, only two sets are possible here. Out
of these, the set corresponding to $b_1 =
 b^{\prime}_1 = (1-\lambda)/2$ alone, will give a positive value for
$n$. Thus,  substituting these values of residues
in \eqref{3es17a},
one obtains the  expression for the  energy eigenvalue as
\begin{equation}
E_n = \frac{\pi^2 }{2ma^2}\left(\frac{1}{2} + n
+\sqrt{\frac{1}{4}-\frac{2mV_0a^2}{\pi^2 \hbar^2}}\right)^2, \label{3e17}
\end{equation}
where $n$ can take positive integral values. Proceeding as above
one obtains the Jacobi differential equation in terms of $t$ for
the polynomial part :
\begin{equation}
(1-t^2)P^{\prime\prime}_n(t) -2t(-n-s+\frac{1}{2})P^{\prime}_n(t) -n(n
  +2s)P_n(t)= 0.   \label{3es19}
\end{equation}
The expression for the wave function is then given by
\begin{equation}
\psi(y) = (y^2 +1)^{-\frac{\lambda}{2}}P^{s_1, s_2}_n(-iy),\label{3es20a}
\end{equation}
where $s_1 = s_2 = -n-s-1/2$.

Hence, as pointed out in the beginning, the two different sectors
of the Hamiltonian, as a function of the coupling parameter and
the eigenvalues emerge from general principles of QHJ formalism,
relying on the singularity structure of the QMF function. The wave
functions corresponding to the definite eigenvalues  are obtained
at the end, which match with the known results \cite{kh}.

{\bf Conclusions}\\
      We have mapped the entire Scarf problem, containing the
bound state and energy bands, to the zero energy sector of a
different Hamiltonian, which is quasi-exactly solvable. This was
achieved through point canonical transformations which led to the
redistribution of singularities in the complex domain. The
singularity structure of this new Hamiltonian is transparent
enough to clearly isolate two different regimes, as a function of
the coupling constant. When related to the original problem, they
turn out to represent discrete levels and the band edges. It will
be interesting to carefully analyze the equilibrium structure of
the classical electrostatics problem, associated with the QES
system, which leads to both bound states and band structure in the
quantum domain. In light of the current interest in periodic
potentials in BEC and photonic crystals, we hope the quantum
Hamilton-Jacobi based treatment presented here is not only
illuminating, but may also lead to development of new perturbative
treatments for non-exactly solvable problems.

{\bf Acknowledgements}\\
We are thankful to Dr.J. Banerji for a careful reading of the
manuscript and R. Atre for his help during the course of this
work.


\begin{thebibliography}{999}
\bibitem{br} J. C. Bronski. L. D. Carr, B. Deconinck and J. N.
Kutz, {\it Phys. Rev. Lett}. {\bf 86}, 1402 (2001).

\bibitem{fo} C. Fort, F. S. Cataliotti, L. Fallani, P.
Maddaloni and M. Inguscio, {\it Phys. Rev. Lett}, {\bf 90}, 140405
(2003).

\bibitem{dim} D. G. Angelakis, M. F. Santos, V. Yannapapas
and A. Ekert, preprint: quant - ph/0410189.

\bibitem{jo} J. D. Joannopoulos, R. D. Meade and J. N. Winn,
{\it Photonic crystals} (Princeton University Press, 1995) and
references therein.

\bibitem{fs} F. Szmulowicz, {\it Am. J. Phys.} {\bf 65}(10) 1009
(1997).

\bibitem{fs1}F. Szmulowicz, {\it Am. J. Phys.} {\bf 72}(11) 1392
(2004).

\bibitem{na} M. Greiner, O. Mandel, T. Esslinger, T. W.
H\"ansch and I. Bloch, {\it Nature}, {\bf 415}, 39 (2002).

\bibitem{kit} C. Kittel, {\it Introduction to Solid State
Physics} (Seventh edition, Wiley Eastern Limited, New Delhi,
1995).

\bibitem{3scf} F. L. Scarf, {\it Phys. Rev}, {\bf 112}, 1137
  (1958) and references therein.

\bibitem{3li} H. Li and D. Kusnezov, {\it Phys. Rev. Lett.} {\bf
  83}, 1283 (1999).

\bibitem{lea}  R. A. Leacock and M. J. Padgett,
  {\it Phys. Rev. Lett}, {\bf 50}, 3 (1983); Phys. Rev. D
 {\bf 28}, 2491 (1983).

\bibitem{wkb} R. S. Bhalla, A. K. Kapoor and P. K. Panigrahi  {\it
    Phys. Rev. A}, {\bf 54}, 951 (1994).

\bibitem{d2} P. A. M. Dirac,
 {\it Rev. Mod. Phys.} {\bf 17}, 195 (1945) ; P. A. M. Dirac,
 Proc. R. Soc. London, {\bf 113A}, 621 (1927).

\bibitem{jor} P. Jordan, {\it Z. Phys.} {\bf 38}, 513 (1926).

\bibitem{sw} J. Schwinger, {\it Quantum Electrodynamics} (Dover
 Publications, Inc. New York, 1958).

\bibitem{3sree} S. Sree Ranjani, K. G. Geojo, A. K Kapoor and
  P. K. Panigrahi, {\it Mod. Phys. Lett. A.} {\bf
  19}, No. {\bf 19}, 1457 (2004);
  preprint quant - ph/0211168.

\bibitem{geo} K. G. Geogo, S. Sree Ranjani and A. K. Kapoor, J. Phys
  A : Math. Gen. {\bf 36}, 4591 (2003); quant - ph/0207036.

\bibitem{geoth}  K. G. Geojo, {\it Quantum Hamilton - Jacobi
  study of wave functions and energy spectrum of solvable and quasi -
  exactly solvable models}, {\it Ph. D. thesis} submitted to University of
  Hyderabad (2004).

\bibitem{akk} S. Sree Ranjani, A. K. Kapoor and P. K. Panigrahi,
  to be published in {\it Mod. Phys. Lett. A.} {\bf 19} No. {\bf
  27},  2047 (2004); preprint quant  - ph/0312041.

\bibitem{ppqes}  S. Sree Ranjani, A. K. Kapoor and P. K. Panigrahi;
  preprint quant - ph/0403196.

\bibitem{sreeth} S. Sree Ranjani {\it Quantum Hamilton - Jacobi
  Solution for spectra of several one dimensional potentials with
  special properties}, {\it Ph. D. thesis}, submitted to the University
  of Hyderabad (2004).

\bibitem{pt}  S. Sree Ranjani, A. K. Kapoor and P. K. Panigrahi;
  preprint quant - ph/0403054.


\bibitem{kh} F. Cooper, A. Khare  and U. Sukhatme, {\it
  Supersymmetry in Quantum Mechanics} (World Scientific, Singapore,
  2001)
\end{thebibliography}
\end{document}